\documentclass[journal=nalefd,manuscript=article]{achemso}

\usepackage[version=3]{mhchem} 
\usepackage[T1]{fontenc}       
\usepackage{graphicx}

\author{Zhili Zhu}

\altaffiliation{Contributed equally to this work}
\author{Xiaolin Cai}
\altaffiliation{Contributed equally to this work}

\author{Chunyao Niu}
\author{Chongze Wang}
\author{Qiang Sun}
\affiliation{International Laboratory for Quantum Functional Materials of Henan, and School of Physics and Engineering,
Zhengzhou University, Zhengzhou 450001, China}
\author{Xiaoyu Han}
\author{Zhengxiao Guo}
\email{z.x.guo@ucl.ac.uk}
\affiliation{Department of Chemistry, University Colledge London, London WC1E 6BT, United Kingdom}
\author{Yu Jia}
\email{jiayu@zzu.edu.cn}
\affiliation{International Laboratory for Quantum Functional Materials of Henan, and School of Physics and Engineering, Zhengzhou University, Zhengzhou 450001, China}
\alsoaffiliation{Center for Advanced Analysis and Computional Science, Zhengzhou University, Zhengzhou 450001, China}
\alsoaffiliation{Key Laboratory for Special Functional Materials(MOE), Henan University, Kaifeng, China}

\title[An \textsf{achemso} demo]
  {Tellurene-a monolayer of tellurium from first-principles prediction}


\begin{document}

\begin{abstract}
A two dimensional (2D) Group-VI Te monolayer, tellurene, is predicted by using first-principles calculations, which consists of  four-membered planner and  six-membered chair-like rings arranged alternately in a 2D lattice. The phonon spectra calculations, combined with \emph{ab initio} molecular dynamics (MD) simulations, demonstrate that tellurene is kinetically stable. The tellurene shows a desirable direct band gap of 1.04 eV and its band structure can be effectively tuned by strain. The effective mass calculations imply that tellurene should also exhibit a relatively high carrier mobility, e.g. compared with MoS$_2$. The significant direct band gap and the high carrier mobility imply that tellurene is a very promising candidate for a new generation of nanoelectronic devices.

\end{abstract}

\section{Introduction}
Research in 2D materials, as inspired by the development of graphene, has experienced an explosive increase in recent years due to their rather unique and exceptional properties with promising applications in electronic, photonic, energy and environmental devices. The 2D group-IV materials including silicene,\cite{feng2012evidence} germanene \cite{davila2014germanene} and stanene \cite{zhu2015epitaxial} have been realized experimentally after graphene. For group-V elements, few-layer black phosphorus, named phosphorene, \cite{li2014black,xia2014rediscovering,zhu2014magnetism} has also been successfully fabricated by exfoliation, which exhibits prominent properties such as high carrier mobility and high on/off ratio. Very recently, the novel 2D group-III material of borophene has been fabricated successfully.\cite{mannix2015synthesis,feng2015experimental}  Beside the allotropes of single element in 2D family, the 2D transition metal dichalcogenides, such as MoS$_2$, \cite{lee2010anomalous} MoSe$_2$, WS$_2$ \cite{elias2013controlled,gutierrez2012extraordinary} and WSe$_2$ \cite{huang2013large,liu2013role}, have been synthesized and attracted both experimental and theoretical interests because of their relatively large and direct band gap as well as good  carrier mobilities. To the best of our knowledge, except these binary compounds containing chalcogen, 2D monolayered structures of simple group VI elements has not been reported before.

The group-VI elements possess a valence configuration of $ns^{2}$$np^{4}$, and their stable structures show systematic change from diatomic molecules (O) through rings, chains and helices (S, Se, Te) to the only simple cubic lattice found in one element (Po). The most stable structure of Te at the atmospheric pressure is a trigonal one, which consists of helical chains parallel to the \emph {c}-axis,\cite{Donohue1974} and each atom covalently bonds to two neighbors within a chain while the interchain bonding is relatively weak. This inherent anisotropy makes it an ideal candidate for the generation of 1D nanostructures. For example, 1D Te nanocrystals, including Te nanotubes, nanorods, nanowires, and nanobelts, have been synthesized through different routes.\cite{mo2002controlled,zhou2006general,lu2004biomolecule,li2004synthesis,zhang20071d,song2008superlong}

Despite the strong tendency to grow into one-dimensional nanoarchitecture, it is also possible for Te to form 2D structure. Recently, the hexagonal Te nanoplates have been successfully realized on flexible mica sheets via van der Waals epitaxy.\cite{wang2014van} However, the Te hexagonal nanoplate has a thickness about 32 nm, which is different from the atomically thinned 2D crystal. Although one monolayer of Te  has been tried to deposite on the CdTe (111) surface with the molecular beam epitaxy method,\cite{ren2016effective} the crystal structure of 2D monolayer Te could not be determined because of the strong covalent bonding between the adsorbed Te atoms and substrate. Therefore, whether the 2D monolayer structure can be realized in the scope of chalcogen, just as succeeded in carbon group and pnictogens, is still an open question.

Here, we investigated the possibilities of 2D materials for chalcogen based on first-principles calculations, and a novel 2D semiconductor, namely tellurene (Te monolayer), is predicted. It is composed of planner four-membered and chair-like six-membered rings alternately arranged in a 2D lattice. The monolayer tellurene shows a direct band gap of 1.04 eV. Interestingly, both the band gap and transport properties of tellurene can be tuned by strains, which make it promising in applications such as electronic and optoelectronic devices.

\section{Results and discussion}

As the reference, the optimized structure of trigonal tellurium (t-Te) is a monoclinic phase with space group P3$_1$21, which consists of helical chains arranged in a hexagonal array along \emph{c} direction, shown in Figure 1a. Figure 1b presents the proposed structure of tellurene, which is composed of planner four-membered and chair-like six-membered rings arranged alternately in a 2D lattice. The optimized unit cell is spanned by lattice vectors $\vec{a}$ and $\vec{b}$, with a = 5.49 {\AA} and b = 4.17 {\AA}, respectively. In tellurene, Te atom favors threefold and fourfold coordinations instead of the twofold coordination in its bulk counterpart. The short and long bonds are 2.75 and 3.01 {\AA}, respectively. The van der Waals radii of the Te atom is about  2.36 {\AA} \cite{batsanov2001van} which indicates that covalent bonds are dominant in tellurene. As shown in Figure 1c, the calculated charge density difference illustrates that, besides the strong covalent bonds between two Te atoms with the coordination number (CN) of three, the Te atoms with CN = 4 also bond to its four neighbor Te atoms. The vanished strong interlayer van der Waals interactions in t-Te may play an important role in the structure transition from bulk helical chains to the hypervalent monolayer structure. The atoms of tellurene are much closer together than those of bulk phase, due to the vanished van der Waals interactions.

\begin{figure}[htbp]
\centering
\begin{minipage}[b]{0.9\textwidth}
\includegraphics[width=12cm]{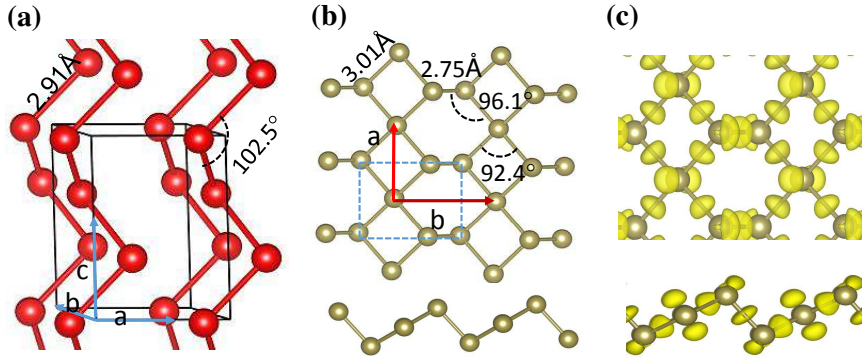}
\end{minipage}
\caption{(a) Schematics of t-Te. (b) Top and side views of monolayered Te. The rectangle represents the unit cell of tellurene in (b) panel. (c) Top and side views of charge density differences (only positive charge density is shown, isosurface set to 0.004 e/$a_{0}^{3}$). \label{fig:Figure1}}
\end{figure}

To examine the relative stability of tellurene, the formation energy is studied by calculation of the cohesive energy difference $\delta$$E_c$ = $E_c$(3D)- $E_c$(2D), where $E_c$(2D) and $E_c$(3D) are the cohesive energies in the 2D tellurene and 3D structure of t-Te, respectively.  In general, this is the energy cost to synthesize the single-layer material from its three-dimensional bulk counterpart. The lower formation energy indicates that the candidate single-layer material is more stable. The calculated $\delta$$E_c$ of tellunene is 0.23 eV/atom, much smaller than the formation energies of silicene (0.76 eV/atom) and germanene (0.99 eV/atom) having been experimentally fabricated, \cite{cahangirov2009two,feng2012evidence,davila2014germanene} which indicates that tellurene is also expected to be synthesized in the experiment.

Actually, it is very important to examine the stability and feasibility in experiment for new 2D crystals. Phonon calculations, reflecting whether there are soft modes, provide a criterion to judge the structure stability.\cite{li2014be2c} Hence, the phonon spectrum is examined, shown in Figure 2 (left panel). It is clearly revealed  no soft phonon modes in the computed phonon dispersion spectrum of tellurene, predicting that tellurene is kinetically stable. The displacement patterns corresponding to the six optical modes at $\Gamma$ point are plotted in right panels in Figure 2, which illustrate the different characteristics of bond stretching and bending . Among these displacement patterns, (a) and (d) correspond to the LO modes at $\Gamma$ point, (c) and (e) correspond to the TO modes, and (b) and (f) correspond to the two ZO modes, respectively. Mode (a) results in a diametric expansion/contraction in the basal plane, and the harder Mode (f) corresponds to rotational motion.

\begin{figure}[htbp]
\centering
\begin{minipage}[b]{0.9\textwidth}
\includegraphics[width=12cm]{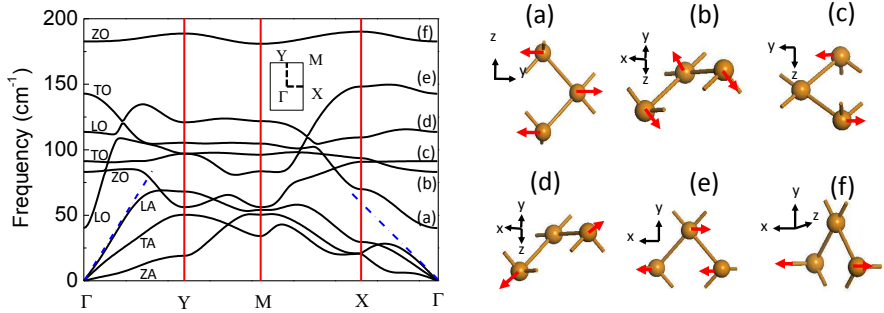}
\end{minipage}
\caption{Phonon band dispersions of tellurene (left panel). ZA mode is the out-of-plane transverse acoustic mode; TA and LA modes are the in-plane transverse and longitudinal acoustic modes; TO and LO modes are the in-plane transverse and longitudinal optical modes; ZO mode is the out-of-plane transverse optical mode. The slope of the dashed lines along the longitudinal acoustic branches near $\Gamma$ point corresponds to the speed of sound and the in-plane stiffness. Right panels show the displacement patterns of six optical modes at $\Gamma$ point shown in the vibrational band structure. \label{fig:Figure2}}
\centering
\end{figure}

To further understand the stability and structural rigidity of tellurene, the slopes of the longitudinal acoustic branches near $\Gamma$ point, which correspond to the speed of sound and reveal the in-plane stiffness, are calculated. As seen in Figure 2, the speed of sound in tellurene is $v_s^{\Gamma-Y} = 3.19\;Km/s$ along the $\Gamma$-Y direction, and $v_s^{\Gamma-X} = 1.76\;Km/s$ along the $\Gamma$-X direction. The speed of sound along the $\Gamma$-Y direction is faster than that along $\Gamma$-X, which reflects an anisotropy in the elastic constants of tellurene. The lower rigidity along the $\Gamma$-X direction, corresponding to the $\vec{a}$ direction in Figure 1b, reflects the fact that compression along the armchair direction requires primarily bond bending, which is lower in energy cost than bond stretching. The anisotropic characters of the elastic constants in tellurene are similar to that in phosphonene.\cite{zhu2014semiconducting} Furthermore, the calculated flexural rigidity D also provides a quantitative explanation for the anisotropy in the elastic constants of tellurene, which can be extracted from total energy of nanotube with different diameter. The calculation details are presented in the Support Information (Figure S1). We have estimated $D = 1.36\;eV$ along the armchair direction and $D = 2.60\;eV$ along the zigzag direction for tellurene, respectively.

\begin{figure}
\centering
\begin{minipage}[b]{1.0\textwidth}
\includegraphics[width=14cm]{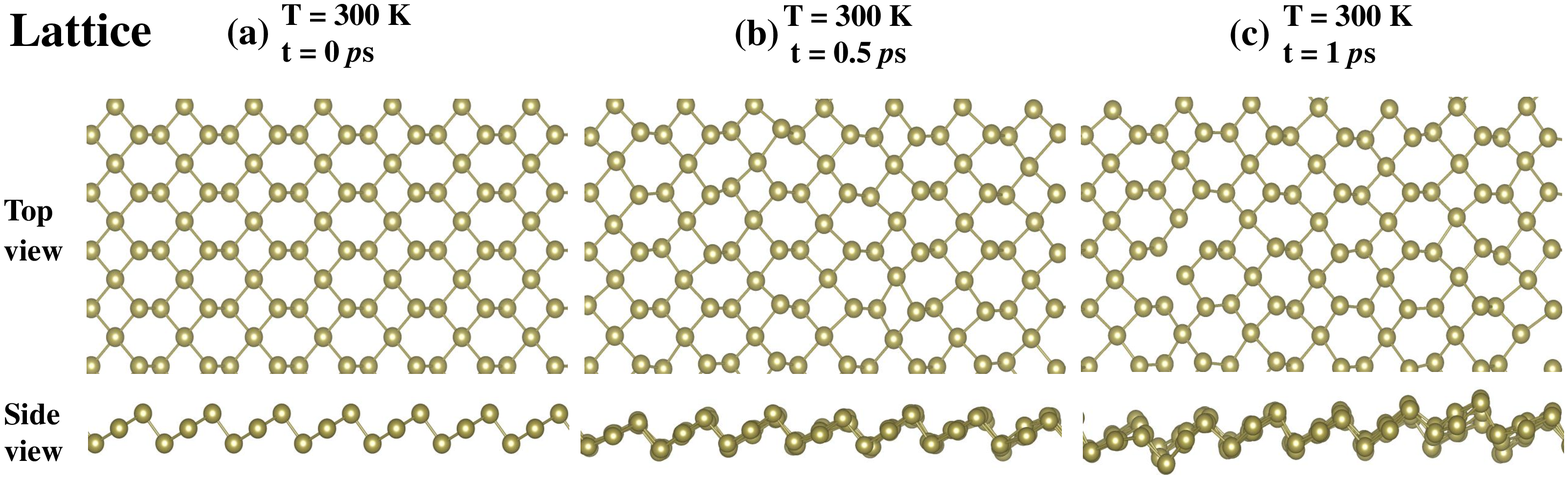}
\end{minipage}
\caption{(a-c) Snapshots of the \emph{ab initio} MD simulations for monolayer tellurene at T = 300 K corresponding to the annealing time of 0, 0.5 and 1 \emph {p}s, respectively. \label{fig:Figure3}}
\centering
\end{figure}

To further clarify the stability of tellurene at elevated temperatures, we performe the \emph{ab initio} MD calculations of the monolayer as well as flakes. The temperatures are kept at 300 and 1000 K, respectively, covering time periods up to 1 \emph {p}s with a time step of 1 \emph {f}s. The high rigidity of a free-standing tellurene monolayer is confirmed by the MD calculations at nonzero temperatures. As shown in Figure 3, the infinite monolayer of tellurene exhibits slight changes at room temperature, revealing the high structural rigidity. The detailed results are presented in the Supplemental Information (Figure S2 and movies).

\begin{figure}[htbp]
\centering
\begin{minipage}[b]{1.0\textwidth}
\includegraphics[width=11cm]{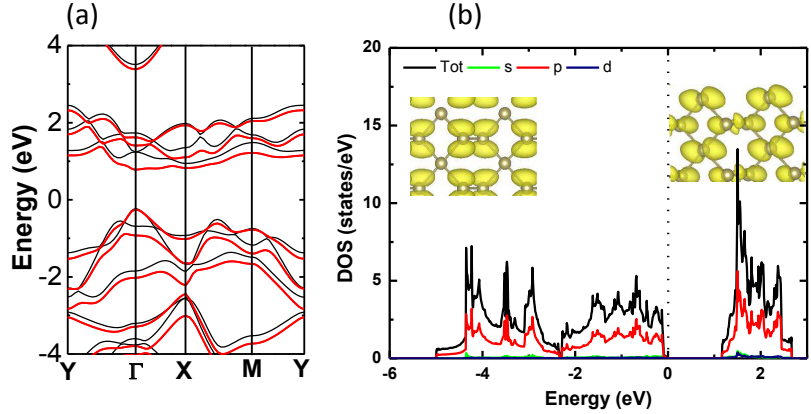}
\end{minipage}
\caption{(a) Band structures of tellurene without (black lines) and with (red lines) spin-orbital coupling, respectively. (b) Total and partial density of states (DOS) of tellurene. The insets denote the calculated partial charges of valence band top (VBM) and conduction band bottom (CBM) (isosurface set to 0.004 e/$a_{0}^{3}$). The vertical dash line indicates the Fermi level. \label{fig:Figure4}}
\centering
\end{figure}

The band structures and the electron density of states are also calculated to explore the electronic properties of tellurene, shown in Figure 4. It is distinctly found that the spin-orbital coupling (SOC) has a significant influence on the electronic structure of tellurene. As shown in Figure 4a, the band structure shows an indirect band gap without SOC, but exhibits a direct band gap of 1.04 eV with SOC. Figure 4b presents the calculated total and partial DOS of tellurene. It is obviously manifested that the DOS near the Fermi level is predominantly \emph{p} like states with almost negligible contributions of \emph{s} and \emph{d} electrons. To gain further insights into the electronic structure, the spatial distributions of the valence band maximum (VBM) and the conduction band minimum (CBM) of tellurene are also examined (Insets of Figure 4b). Both the VBM and CBM originate mainly from the \emph{p} atomic orbitals, and the VBM is bonding-like along the armchair direction while CBM is antibonding-like state.

\begin{figure}[htbp]
\centering
\begin{minipage}[b]{1.0\textwidth}
\includegraphics[width=10cm]{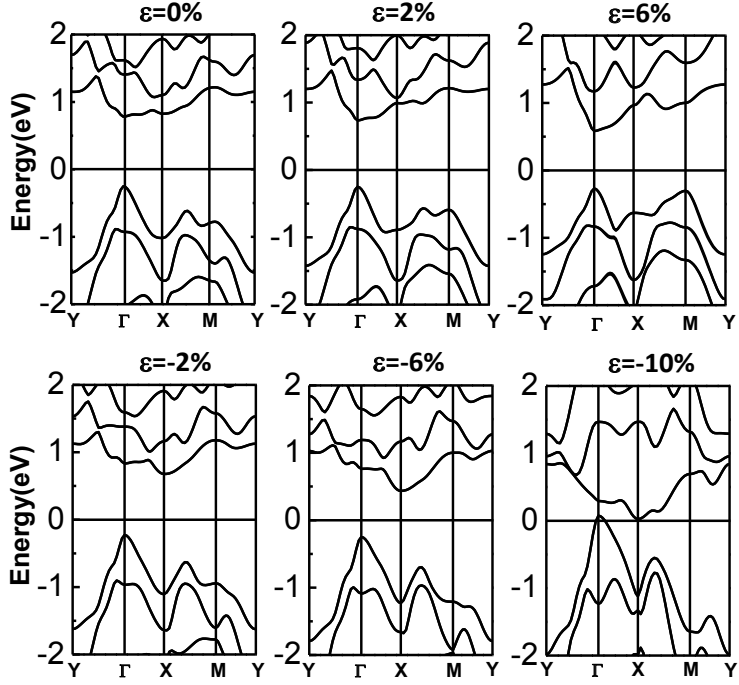}
\end{minipage}
\caption{Band structures of tellurene under different biaxial strains (from $\epsilon$ = -10\% to 6\%) with SOC. \label{fig:Figure5}}
\centering
\end{figure}

It has been shown that applying external strain is an effective way to tune the electronic properties of 2D materials.\cite{morgan2015compressive, zhang2015magnetic,  han2014strain} The band gap of tellurene  also depends sensitively on the applied in-layer strain. Figure 5 presents the band structures of tellurene under the biaxial strains from -10\% to 6\%. It is found that CBM shifts down gradually towards the Fermi Level as the  tensile strain increases from 0\% to 6\%, whereas the energy of VBM almost does not change, resulting in the reduced band gap, and the band gap reduced to 0.86 eV under the tensile strain of 6\%. This manifests that the band gap of stretched tellurene is dominated by the CBM.  Nevertheless, the stretched tellurene still retains the character of direct band gap. On the other hand, the monolayer tellurene undergoes a direct-to-indirect band-gap transition under the compressed strain. It is found that a relatively small critical strain about -2\% can induce a direct-to-indirect band gap transition. The CBM shifts from $\Gamma$ to X point under the compressed strain, which rapidly transforms tellurene into an indirect band gap semiconductor. In tellurene, the bonds among Te atoms may have the dual characteristics of metal and covalence, and there exists a competition between them in the strained stages. The covalent bonding may be dominant under the stretched strain, which will decrease the bonding between the \emph{p} orbitals, leading to the separation of bonding and antibonding states getting smaller near the Fermi level. On the other hand, the metal bonding characteristic may be critical when the tellurene is compressed. The overlaps between \emph{p} orbitals of Te atoms will be strengthened as the compressed strain increases, which improves the power of charge transportation. Meanwhile, the band gap becomes small and presents a metallic character under $\epsilon$ = -10\%. 

\begin{figure}[htbp]
\centering
\begin{minipage}[b]{1.0\textwidth}
\includegraphics[width=10cm]{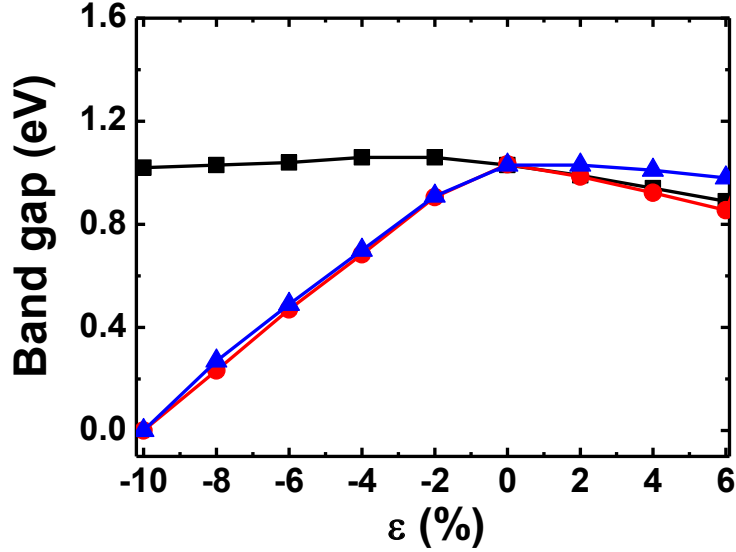}
\end{minipage}
\caption{Variations of band gap with the biaxial strain (red line with circles), uniaxial strains along armchair ( black line with squares) and zigzag (blue line with triangles) directions, respectively. \label{fig:Figure6}}
\centering
\end{figure}

In order to better understand the strain effects on the electronic properties of tellurene, uniaxial strains along armchair and zigzag directions of the lattice are also applied, respectively (Figure S3 and Figure S4). The results suggest that the strain effects of tellurene exhibit very strong anisotropy. As shown in Figure 6, the compressed strain along the zigzag direction critically influences the electronic structure, which modulates the band gap of tellurene, similar to that of biaxial strain. However, the band gap does not change much  along armchair direction under the compressed strain from 0\% to -10\%. Moreover, the tensile strain along the armchair direction shows more significant effect on tuning the band structure of tellurene than that along the zigzag direction from $\epsilon$ = 0\% to 6\%.

One of the main interests in 2D semiconductors is the observed high mobility of carriers such as graphene and phosphorene. In order to provide further insights into the charge transport properties of tellurene, the effective masses of electrons at the CBM (\emph{m}$_{e}^{*}$) and of holes at the VBM (\emph{m}$_{h}^{*}$) are calculated using
\[m^* = \frac{h^2}{4\pi^2} (\frac{\partial^2 E}{\partial k^2})^{-1},\]
where \emph{E} and \emph{k} correspond to the energy and the reciprocal lattice vector along the axis. In solid theory, it is thought that the charge carrier mobility is inversely proportional to the effective mass if there is no change in the time scale for quasi particle scattering. Based on this, we can simply define the mobility $\mu$ as

\[\mu = \frac{q}{m^*} \tau.\]
According to the above formula, changes in effective mass can influence charge carrier transport properties. Under zero strain, the effective electron masses are \emph{m}$_{ex}^{*}$= 0.83 \emph{m}$_0$ and \emph{m}$_{ey}^{*}$=0.19 \emph{m}$_0$, and the effective hole masses are \emph{m}$_{hx}^{*}$= 0.39 \emph{m}$_0$ and \emph{m}$_{hy}^{*}$=0.12 \emph{m}$_0$, respectively. Here, \emph{m}$_0$ is the mass of electron. The effective mass of hole is lower than that of the electron which indicates that the carrier mobility is dominated by holes in intrinsic monolayer tellurene. Compare with the effective electron mass \emph{m}$_{e}^{*}$= 0.48 \emph{m}$_0$ of monolayer MoS$_2$ \cite{kaasbjerg2012phonon}, it is likely that tellurene will exhibit a higher carrier mobility than monolayer MoS$_2$.

\begin{figure}[htp]
\centering
\includegraphics[width=10cm]{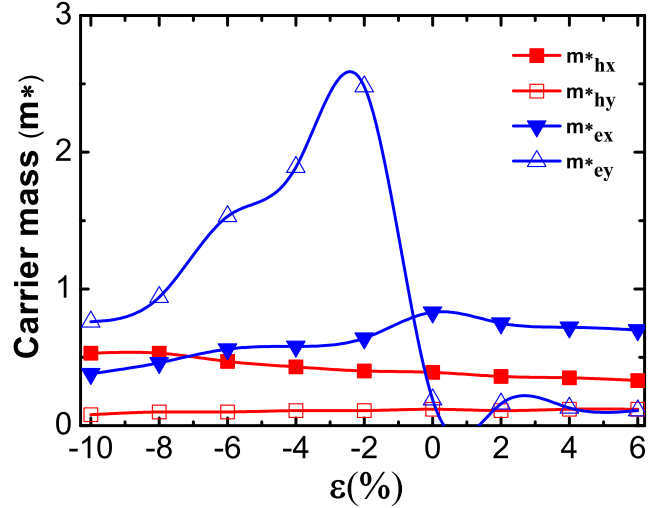}
\caption{Variations of hole and electronic effective masses with respect to the biaxial strain for tellurene. \label{fig:Figure7}}
\centering
\end{figure}

The variations of effective masses of hole and electron with the biaxial strain are also studied in tellurene, presented in Figure 7. The results suggest that the compressed strain significantly affects the electron effective mass, i.e. strongly influences the electron transport of tellurene. The \emph{m}$_{ey}^{*}$ shows a significant jump at the maximum values of 2.48 \emph{m}$_0$ under the compressed strain of -2\%, and subsequently decreases with the strain from -2\% to -10\%. This result can be also reflected in the band structure variations with strain, in which the CBM shifts from $\Gamma$  to X resulting in the transition from direct to indirect band gap semiconductor. On the other hand, the \emph{m}$_{ex}^{*}$ decreases from 0.83 \emph{m}$_0$ under $\epsilon$ = 0\% to 0.36 \emph{m}$_0$ under $\epsilon$ = -10\%. However, the tensile strain does not modify the effective masses much between $\epsilon$ = 0\% and 6\%.

The electronic and hole effective masses under the uniaxial strains are also examined (Figure S5 and Figure S6). Although it slightly modifies the band gap of tellurene, the compressed strain along the armchair direction critically influences the electronic effective masses. The \emph{m}$_{ey}^{*}$ decreases from 3.20 \emph{m}$_0$ under $\epsilon$ = -2\% to 1.73 \emph{m}$_0$ under $\epsilon$ = -10\%. In addition, the \emph{m}$_{hx}^{*}$ can also be effectively tuned from 0.42 \emph{m}$_0$ to 0.75 \emph{m}$_0$ between $\epsilon$ = -2\% and -10\%. On the other hand, it is found that the compressed strain  from $\epsilon$ = -4\% to -10\% along the zigzag direction does not cause much change in the \emph{m}$_{ey}^{*}$. The tensile strain leads to an increase of \emph{m}$_{ex}^{*}$ from 0.63 \emph{m}$_0$ under $\epsilon$ = 0\% to 1.07 \emph{m}$_0$ under $\epsilon$ = 6\%. Surprisingly, the \emph{m}$_{ex}^{*}$ undergoes a significant jump from 0.51 \emph{m}$_0$ under $\epsilon$ = -8\% to 1.51 \emph{m}$_0$ under $\epsilon$ = -10\%.

From the above analysis of the behavior of the effective mass, it is likely that the proposed tellurene may exhibit a higher carrier mobility than MoS$_2$. Moreover, the transport properties can be modified by both biaxial and uniaxial strains. The combination of a significant band gap and a high carrier mobility implies that tellurene is a highly promising candidate for a new generation of nanoelectronic devices.

In summary, a 2D group-VI material, tellurene, has been predicted for the first time. It is composed of planner four-membered and chair-like six-membered rings  arranged alternately in a 2D lattice. The monolayer tellurene shows a direct band gap of 1.04 eV. In addition, both the band gap and transport properties of tellurene can be tuned by external strain. All these characteristics make tellurene a highly promising candidate for applications in electronic, photonic and optoelectronic devices.

\section{Methods}

All calculations are performed within the density functional theory (DFT) using the Vienna  \emph{ab initio} Simulation Package (VASP).\cite{kresse1996efficient} The projected augmented wave (PAW) potentials \cite{blochl1994projector,kresse1999ultrasoft} are adopted to treat the core electrons, and the Perdew-Burke-Ernzerhof (PBE) exchange-correlation functional \cite{perdew1996generalized} is employed. The unit cell including a vacuum of 15 {\AA} is used to simulate the 2D monlayer crystal. For Brillouin zone (BZ) integration, the Monkhorst-Pack scheme \cite{monkhorst1976special} with a 15 $\times$ 15 $\times$ 1 k-point grid is used. The kinetic energy cutoff for the plane wave basis set is chosen to be 500 eV. All structures are fully relaxed until the total energies are converged up to $10^{-4}$ eV and the Hellmann-Feynman forces are less than 0.02 eV/{\AA}.
The phonon calculations employ a supercell approach, as implemented in the Phonopy code.\cite{togo2008first} In the  \emph{ab initio} MD simulations, the temperatures are kept at 300 and 1000 K for 1 \emph{p}s with a time step of 1 \emph{f}s. The van der Waals effect is treated using the DFT-D2 method of Grimme.\cite{zacharia2004interlayer}

\begin{suppinfo}

Flexural rigidity (D) calculations of tellurene,  \emph{ab initio} MD simulations of infinite monolayer as well as a free-standing finite flake of tellurene at high temperatures, uniaxial strain effects on the band structures, effective masses modulated by uniaxial strains.

\end{suppinfo}

\begin{acknowledgement}

The authors gratefully acknowledge financial support from the National Natural Science Foundation of China (No. 11274280 and 11504332 ), the National Basic Research Program of China (No. 2012CB921300), the Natural Science Foundation of Henan Province of China (No.152300410049) and the Key Scientific Research Project of Henan Province (No. 16A140019). The calculations are performed on the High Performance Clusters of Zhengzhou University. ZXG also acknowledge the support of the UK EPSRC (No. EP/K021192/1).

\end{acknowledgement}

\providecommand{\latin}[1]{#1}
\providecommand*\mcitethebibliography{\thebibliography}
\csname @ifundefined\endcsname{endmcitethebibliography}
  {\let\endmcitethebibliography\endthebibliography}{}


\begin{mcitethebibliography}{38}
\providecommand*\natexlab[1]{#1}
\providecommand*\mciteSetBstSublistMode[1]{}
\providecommand*\mciteSetBstMaxWidthForm[2]{}
\providecommand*\mciteBstWouldAddEndPuncttrue
  {\def\EndOfBibitem{\unskip.}}
\providecommand*\mciteBstWouldAddEndPunctfalse
  {\let\EndOfBibitem\relax}
\providecommand*\mciteSetBstMidEndSepPunct[3]{}
\providecommand*\mciteSetBstSublistLabelBeginEnd[3]{}
\providecommand*\EndOfBibitem{}
\mciteSetBstSublistMode{f}
\mciteSetBstMaxWidthForm{subitem}{(\alph{mcitesubitemcount})}
\mciteSetBstSublistLabelBeginEnd
  {\mcitemaxwidthsubitemform\space}
  {\relax}
  {\relax}

\bibitem[Feng \latin{et~al.}(2012)Feng, Ding, Meng, Yao, He, Cheng, Chen, and
  Wu]{feng2012evidence}
Feng,~B.; Ding,~Z.; Meng,~S.; Yao,~Y.; He,~X.; Cheng,~P.; Chen,~L.; Wu,~K.
  \emph{Nano letters} \textbf{2012}, \emph{12}, 3507--3511\relax
\mciteBstWouldAddEndPuncttrue
\mciteSetBstMidEndSepPunct{\mcitedefaultmidpunct}
{\mcitedefaultendpunct}{\mcitedefaultseppunct}\relax
\EndOfBibitem
\bibitem[D{\'a}vila \latin{et~al.}(2014)D{\'a}vila, Xian, Cahangirov, Rubio,
  and Le~Lay]{davila2014germanene}
D{\'a}vila,~M.; Xian,~L.; Cahangirov,~S.; Rubio,~A.; Le~Lay,~G. \emph{New
  Journal of Physics} \textbf{2014}, \emph{16}, 095002\relax
\mciteBstWouldAddEndPuncttrue
\mciteSetBstMidEndSepPunct{\mcitedefaultmidpunct}
{\mcitedefaultendpunct}{\mcitedefaultseppunct}\relax
\EndOfBibitem
\bibitem[Zhu \latin{et~al.}(2015)Zhu, Chen, Xu, Gao, Guan, Liu, Qian, Zhang,
  and Jia]{zhu2015epitaxial}
Zhu,~F.-f.; Chen,~W.-j.; Xu,~Y.; Gao,~C.-l.; Guan,~D.-d.; Liu,~C.-h.; Qian,~D.;
  Zhang,~S.-C.; Jia,~J.-f. \emph{Nature materials} \textbf{2015}, \emph{14},
  1020--1025\relax
\mciteBstWouldAddEndPuncttrue
\mciteSetBstMidEndSepPunct{\mcitedefaultmidpunct}
{\mcitedefaultendpunct}{\mcitedefaultseppunct}\relax
\EndOfBibitem
\bibitem[Li \latin{et~al.}(2014)Li, Yu, Ye, Ge, Ou, Wu, Feng, Chen, and
  Zhang]{li2014black}
Li,~L.; Yu,~Y.; Ye,~G.~J.; Ge,~Q.; Ou,~X.; Wu,~H.; Feng,~D.; Chen,~X.~H.;
  Zhang,~Y. \emph{Nature nanotechnology} \textbf{2014}, \emph{9},
  372--377\relax
\mciteBstWouldAddEndPuncttrue
\mciteSetBstMidEndSepPunct{\mcitedefaultmidpunct}
{\mcitedefaultendpunct}{\mcitedefaultseppunct}\relax
\EndOfBibitem
\bibitem[Xia \latin{et~al.}(2014)Xia, Wang, and Jia]{xia2014rediscovering}
Xia,~F.; Wang,~H.; Jia,~Y. \emph{Nature communications} \textbf{2014},
  \emph{5}\relax
\mciteBstWouldAddEndPuncttrue
\mciteSetBstMidEndSepPunct{\mcitedefaultmidpunct}
{\mcitedefaultendpunct}{\mcitedefaultseppunct}\relax
\EndOfBibitem
\bibitem[Zhu \latin{et~al.}(2014)Zhu, Li, Yu, Chang, Sun, and
  Jia]{zhu2014magnetism}
Zhu,~Z.; Li,~C.; Yu,~W.; Chang,~D.; Sun,~Q.; Jia,~Y. \emph{Applied Physics
  Letters} \textbf{2014}, \emph{105}, 113105\relax
\mciteBstWouldAddEndPuncttrue
\mciteSetBstMidEndSepPunct{\mcitedefaultmidpunct}
{\mcitedefaultendpunct}{\mcitedefaultseppunct}\relax
\EndOfBibitem
\bibitem[Mannix \latin{et~al.}(2015)Mannix, Zhou, Kiraly, Wood, Alducin, Myers,
  Liu, Fisher, Santiago, Guest, \latin{et~al.} others]{mannix2015synthesis}
others,, \latin{et~al.}  \emph{Science} \textbf{2015}, \emph{350},
  1513--1516\relax
\mciteBstWouldAddEndPuncttrue
\mciteSetBstMidEndSepPunct{\mcitedefaultmidpunct}
{\mcitedefaultendpunct}{\mcitedefaultseppunct}\relax
\EndOfBibitem
\bibitem[Feng \latin{et~al.}(2015)Feng, Zhang, Zhong, Li, Li, Li, Cheng, Meng,
  Chen, and Wu]{feng2015experimental}
Feng,~B.; Zhang,~J.; Zhong,~Q.; Li,~W.; Li,~S.; Li,~H.; Cheng,~P.; Meng,~S.;
  Chen,~L.; Wu,~K. \emph{arXiv preprint arXiv:1512.05029} \textbf{2015}, \relax
\mciteBstWouldAddEndPunctfalse
\mciteSetBstMidEndSepPunct{\mcitedefaultmidpunct}
{}{\mcitedefaultseppunct}\relax
\EndOfBibitem
\bibitem[Lee \latin{et~al.}(2010)Lee, Yan, Brus, Heinz, Hone, and
  Ryu]{lee2010anomalous}
Lee,~C.; Yan,~H.; Brus,~L.~E.; Heinz,~T.~F.; Hone,~J.; Ryu,~S. \emph{ACS nano}
  \textbf{2010}, \emph{4}, 2695--2700\relax
\mciteBstWouldAddEndPuncttrue
\mciteSetBstMidEndSepPunct{\mcitedefaultmidpunct}
{\mcitedefaultendpunct}{\mcitedefaultseppunct}\relax
\EndOfBibitem
\bibitem[Elias \latin{et~al.}(2013)Elias, Perea-L{\'o}pez, Castro-Beltr{\'a}n,
  Berkdemir, Lv, Feng, Long, Hayashi, Kim, and Endo]{elias2013controlled}
Elias,~A.~L.; Perea-L{\'o}pez,~N.; Castro-Beltr{\'a}n,~A.; Berkdemir,~A.;
  Lv,~R.; Feng,~S.; Long,~A.~D.; Hayashi,~T.; Kim,~Y.~A.; Endo,~M. \emph{Acs
  Nano} \textbf{2013}, \emph{7}, 5235--5242\relax
\mciteBstWouldAddEndPuncttrue
\mciteSetBstMidEndSepPunct{\mcitedefaultmidpunct}
{\mcitedefaultendpunct}{\mcitedefaultseppunct}\relax
\EndOfBibitem
\bibitem[Guti{\'e}rrez \latin{et~al.}(2012)Guti{\'e}rrez, Perea-L{\'o}pez,
  El{\'\i}as, Berkdemir, Wang, Lv, L{\'o}pez-Ur{\'\i}as, Crespi, Terrones, and
  Terrones]{gutierrez2012extraordinary}
Guti{\'e}rrez,~H.~R.; Perea-L{\'o}pez,~N.; El{\'\i}as,~A.~L.; Berkdemir,~A.;
  Wang,~B.; Lv,~R.; L{\'o}pez-Ur{\'\i}as,~F.; Crespi,~V.~H.; Terrones,~H.;
  Terrones,~M. \emph{Nano letters} \textbf{2012}, \emph{13}, 3447--3454\relax
\mciteBstWouldAddEndPuncttrue
\mciteSetBstMidEndSepPunct{\mcitedefaultmidpunct}
{\mcitedefaultendpunct}{\mcitedefaultseppunct}\relax
\EndOfBibitem
\bibitem[Huang \latin{et~al.}(2013)Huang, Pu, Hsu, Chiu, Juang, Chang, Chang,
  Iwasa, Takenobu, and Li]{huang2013large}
Huang,~J.-K.; Pu,~J.; Hsu,~C.-L.; Chiu,~M.-H.; Juang,~Z.-Y.; Chang,~Y.-H.;
  Chang,~W.-H.; Iwasa,~Y.; Takenobu,~T.; Li,~L.-J. \emph{ACS nano}
  \textbf{2013}, \emph{8}, 923--930\relax
\mciteBstWouldAddEndPuncttrue
\mciteSetBstMidEndSepPunct{\mcitedefaultmidpunct}
{\mcitedefaultendpunct}{\mcitedefaultseppunct}\relax
\EndOfBibitem
\bibitem[Liu \latin{et~al.}(2013)Liu, Kang, Sarkar, Khatami, Jena, and
  Banerjee]{liu2013role}
Liu,~W.; Kang,~J.; Sarkar,~D.; Khatami,~Y.; Jena,~D.; Banerjee,~K. \emph{Nano
  letters} \textbf{2013}, \emph{13}, 1983--1990\relax
\mciteBstWouldAddEndPuncttrue
\mciteSetBstMidEndSepPunct{\mcitedefaultmidpunct}
{\mcitedefaultendpunct}{\mcitedefaultseppunct}\relax
\EndOfBibitem
\bibitem[Donohue(1974)]{Donohue1974}
Donohue,~J. \emph{The structure of the Elements}, 1st ed.; John Wiley and Sons:
  New York, 1974\relax
\mciteBstWouldAddEndPuncttrue
\mciteSetBstMidEndSepPunct{\mcitedefaultmidpunct}
{\mcitedefaultendpunct}{\mcitedefaultseppunct}\relax
\EndOfBibitem
\bibitem[Mo \latin{et~al.}(2002)Mo, Zeng, Liu, Yu, Zhang, and
  Qian]{mo2002controlled}
Mo,~M.; Zeng,~J.; Liu,~X.; Yu,~W.; Zhang,~S.; Qian,~Y. \emph{Advanced
  Materials} \textbf{2002}, \emph{14}, 1658--1662\relax
\mciteBstWouldAddEndPuncttrue
\mciteSetBstMidEndSepPunct{\mcitedefaultmidpunct}
{\mcitedefaultendpunct}{\mcitedefaultseppunct}\relax
\EndOfBibitem
\bibitem[Zhou and Zhu(2006)Zhou, and Zhu]{zhou2006general}
Zhou,~B.; Zhu,~J.-J. \emph{Nanotechnology} \textbf{2006}, \emph{17}, 1763\relax
\mciteBstWouldAddEndPuncttrue
\mciteSetBstMidEndSepPunct{\mcitedefaultmidpunct}
{\mcitedefaultendpunct}{\mcitedefaultseppunct}\relax
\EndOfBibitem
\bibitem[Lu \latin{et~al.}(2004)Lu, Gao, and Komarneni]{lu2004biomolecule}
Lu,~Q.; Gao,~F.; Komarneni,~S. \emph{Advanced Materials} \textbf{2004},
  \emph{16}, 1629--1632\relax
\mciteBstWouldAddEndPuncttrue
\mciteSetBstMidEndSepPunct{\mcitedefaultmidpunct}
{\mcitedefaultendpunct}{\mcitedefaultseppunct}\relax
\EndOfBibitem
\bibitem[Li \latin{et~al.}(2004)Li, Cao, Feng, and Li]{li2004synthesis}
Li,~X.-L.; Cao,~G.-H.; Feng,~C.-M.; Li,~Y.-D. \emph{Journal of Materials
  Chemistry} \textbf{2004}, \emph{14}, 244--247\relax
\mciteBstWouldAddEndPuncttrue
\mciteSetBstMidEndSepPunct{\mcitedefaultmidpunct}
{\mcitedefaultendpunct}{\mcitedefaultseppunct}\relax
\EndOfBibitem
\bibitem[Zhang \latin{et~al.}(2007)Zhang, Hou, Ye, Fu, and Xie]{zhang20071d}
Zhang,~B.; Hou,~W.; Ye,~X.; Fu,~S.; Xie,~Y. \emph{Advanced Functional
  Materials} \textbf{2007}, \emph{17}, 486--492\relax
\mciteBstWouldAddEndPuncttrue
\mciteSetBstMidEndSepPunct{\mcitedefaultmidpunct}
{\mcitedefaultendpunct}{\mcitedefaultseppunct}\relax
\EndOfBibitem
\bibitem[Song \latin{et~al.}(2008)Song, Lin, Zhan, Tian, Liu, and
  Yu]{song2008superlong}
Song,~J.-M.; Lin,~Y.-Z.; Zhan,~Y.-J.; Tian,~Y.-C.; Liu,~G.; Yu,~S.-H.
  \emph{Crystal Growth and Design} \textbf{2008}, \emph{8}, 1902--1908\relax
\mciteBstWouldAddEndPuncttrue
\mciteSetBstMidEndSepPunct{\mcitedefaultmidpunct}
{\mcitedefaultendpunct}{\mcitedefaultseppunct}\relax
\EndOfBibitem
\bibitem[Wang \latin{et~al.}(2014)Wang, Safdar, Xu, Mirza, Wang, and
  He]{wang2014van}
Wang,~Q.; Safdar,~M.; Xu,~K.; Mirza,~M.; Wang,~Z.; He,~J. \emph{ACS nano}
  \textbf{2014}, \emph{8}, 7497--7505\relax
\mciteBstWouldAddEndPuncttrue
\mciteSetBstMidEndSepPunct{\mcitedefaultmidpunct}
{\mcitedefaultendpunct}{\mcitedefaultseppunct}\relax
\EndOfBibitem
\bibitem[Ren \latin{et~al.}(2016)Ren, Fu, Bian, Su, Zhang, Velury, Yukawa,
  Zhang, Wang, Zha, \latin{et~al.} others]{ren2016effective}
others,, \latin{et~al.}  \emph{ACS applied materials \& interfaces}
  \textbf{2016}, \relax
\mciteBstWouldAddEndPunctfalse
\mciteSetBstMidEndSepPunct{\mcitedefaultmidpunct}
{}{\mcitedefaultseppunct}\relax
\EndOfBibitem
\bibitem[Batsanov(2001)]{batsanov2001van}
Batsanov,~S. \emph{Inorganic materials} \textbf{2001}, \emph{37},
  871--885\relax
\mciteBstWouldAddEndPuncttrue
\mciteSetBstMidEndSepPunct{\mcitedefaultmidpunct}
{\mcitedefaultendpunct}{\mcitedefaultseppunct}\relax
\EndOfBibitem
\bibitem[Cahangirov \latin{et~al.}(2009)Cahangirov, Topsakal, Akt{\"u}rk,
  {\c{S}}ahin, and Ciraci]{cahangirov2009two}
Cahangirov,~S.; Topsakal,~M.; Akt{\"u}rk,~E.; {\c{S}}ahin,~H.; Ciraci,~S.
  \emph{Physical review letters} \textbf{2009}, \emph{102}, 236804\relax
\mciteBstWouldAddEndPuncttrue
\mciteSetBstMidEndSepPunct{\mcitedefaultmidpunct}
{\mcitedefaultendpunct}{\mcitedefaultseppunct}\relax
\EndOfBibitem
\bibitem[Li \latin{et~al.}(2014)Li, Liao, and Chen]{li2014be2c}
Li,~Y.; Liao,~Y.; Chen,~Z. \emph{Angewandte Chemie International Edition}
  \textbf{2014}, \emph{53}, 7248--7252\relax
\mciteBstWouldAddEndPuncttrue
\mciteSetBstMidEndSepPunct{\mcitedefaultmidpunct}
{\mcitedefaultendpunct}{\mcitedefaultseppunct}\relax
\EndOfBibitem
\bibitem[Zhu and Tom{\'a}nek(2014)Zhu, and Tom{\'a}nek]{zhu2014semiconducting}
Zhu,~Z.; Tom{\'a}nek,~D. \emph{Physical review letters} \textbf{2014},
  \emph{112}, 176802\relax
\mciteBstWouldAddEndPuncttrue
\mciteSetBstMidEndSepPunct{\mcitedefaultmidpunct}
{\mcitedefaultendpunct}{\mcitedefaultseppunct}\relax
\EndOfBibitem
\bibitem[Morgan~Stewart \latin{et~al.}(2015)Morgan~Stewart, Shevlin, Catlow,
  and Guo]{morgan2015compressive}
Morgan~Stewart,~H.; Shevlin,~S.~A.; Catlow,~C. R.~A.; Guo,~Z.~X. \emph{Nano
  letters} \textbf{2015}, \emph{15}, 2006--2010\relax
\mciteBstWouldAddEndPuncttrue
\mciteSetBstMidEndSepPunct{\mcitedefaultmidpunct}
{\mcitedefaultendpunct}{\mcitedefaultseppunct}\relax
\EndOfBibitem
\bibitem[Zhang \latin{et~al.}(2015)Zhang, Li, Guo, Cho, Su, and
  Jia]{zhang2015magnetic}
Zhang,~S.; Li,~C.; Guo,~Z.~X.; Cho,~J.-H.; Su,~W.-S.; Jia,~Y.
  \emph{Nanotechnology} \textbf{2015}, \emph{26}, 295402\relax
\mciteBstWouldAddEndPuncttrue
\mciteSetBstMidEndSepPunct{\mcitedefaultmidpunct}
{\mcitedefaultendpunct}{\mcitedefaultseppunct}\relax
\EndOfBibitem
\bibitem[Han \latin{et~al.}(2014)Han, Morgan~Stewart, Shevlin, Catlow, and
  Guo]{han2014strain}
Han,~X.; Morgan~Stewart,~H.; Shevlin,~S.~A.; Catlow,~C. R.~A.; Guo,~Z.~X.
  \emph{Nano letters} \textbf{2014}, \emph{14}, 4607--4614\relax
\mciteBstWouldAddEndPuncttrue
\mciteSetBstMidEndSepPunct{\mcitedefaultmidpunct}
{\mcitedefaultendpunct}{\mcitedefaultseppunct}\relax
\EndOfBibitem
\bibitem[Kaasbjerg \latin{et~al.}(2012)Kaasbjerg, Thygesen, and
  Jacobsen]{kaasbjerg2012phonon}
Kaasbjerg,~K.; Thygesen,~K.~S.; Jacobsen,~K.~W. \emph{Physical Review B}
  \textbf{2012}, \emph{85}, 115317\relax
\mciteBstWouldAddEndPuncttrue
\mciteSetBstMidEndSepPunct{\mcitedefaultmidpunct}
{\mcitedefaultendpunct}{\mcitedefaultseppunct}\relax
\EndOfBibitem
\bibitem[Kresse and Furthm{\"u}ller(1996)Kresse, and
  Furthm{\"u}ller]{kresse1996efficient}
Kresse,~G.; Furthm{\"u}ller,~J. \emph{Physical Review B} \textbf{1996},
  \emph{54}, 11169\relax
\mciteBstWouldAddEndPuncttrue
\mciteSetBstMidEndSepPunct{\mcitedefaultmidpunct}
{\mcitedefaultendpunct}{\mcitedefaultseppunct}\relax
\EndOfBibitem
\bibitem[Bl{\"o}chl(1994)]{blochl1994projector}
Bl{\"o}chl,~P.~E. \emph{Physical Review B} \textbf{1994}, \emph{50},
  17953\relax
\mciteBstWouldAddEndPuncttrue
\mciteSetBstMidEndSepPunct{\mcitedefaultmidpunct}
{\mcitedefaultendpunct}{\mcitedefaultseppunct}\relax
\EndOfBibitem
\bibitem[Kresse and Joubert(1999)Kresse, and Joubert]{kresse1999ultrasoft}
Kresse,~G.; Joubert,~D. \emph{Physical Review B} \textbf{1999}, \emph{59},
  1758\relax
\mciteBstWouldAddEndPuncttrue
\mciteSetBstMidEndSepPunct{\mcitedefaultmidpunct}
{\mcitedefaultendpunct}{\mcitedefaultseppunct}\relax
\EndOfBibitem
\bibitem[Perdew \latin{et~al.}(1996)Perdew, Burke, and
  Ernzerhof]{perdew1996generalized}
Perdew,~J.~P.; Burke,~K.; Ernzerhof,~M. \emph{Physical review letters}
  \textbf{1996}, \emph{77}, 3865\relax
\mciteBstWouldAddEndPuncttrue
\mciteSetBstMidEndSepPunct{\mcitedefaultmidpunct}
{\mcitedefaultendpunct}{\mcitedefaultseppunct}\relax
\EndOfBibitem
\bibitem[Monkhorst and Pack(1976)Monkhorst, and Pack]{monkhorst1976special}
Monkhorst,~H.~J.; Pack,~J.~D. \emph{Physical Review B} \textbf{1976},
  \emph{13}, 5188\relax
\mciteBstWouldAddEndPuncttrue
\mciteSetBstMidEndSepPunct{\mcitedefaultmidpunct}
{\mcitedefaultendpunct}{\mcitedefaultseppunct}\relax
\EndOfBibitem
\bibitem[Togo \latin{et~al.}(2008)Togo, Oba, and Tanaka]{togo2008first}
Togo,~A.; Oba,~F.; Tanaka,~I. \emph{Physical Review B} \textbf{2008},
  \emph{78}, 134106\relax
\mciteBstWouldAddEndPuncttrue
\mciteSetBstMidEndSepPunct{\mcitedefaultmidpunct}
{\mcitedefaultendpunct}{\mcitedefaultseppunct}\relax
\EndOfBibitem
\bibitem[Zacharia \latin{et~al.}(2004)Zacharia, Ulbricht, and
  Hertel]{zacharia2004interlayer}
Zacharia,~R.; Ulbricht,~H.; Hertel,~T. \emph{Physical Review B} \textbf{2004},
  \emph{69}, 155406\relax
\mciteBstWouldAddEndPuncttrue
\mciteSetBstMidEndSepPunct{\mcitedefaultmidpunct}
{\mcitedefaultendpunct}{\mcitedefaultseppunct}\relax
\EndOfBibitem
\end{mcitethebibliography}
\end{document}